\documentclass[aps,prl,reprint,superscriptaddress,longbibliography,floatfix]{revtex4-2}
\usepackage[T1]{fontenc}
\usepackage{amsmath,amssymb,bm,graphicx,url}
\usepackage{xcolor,tikz,pgfplots}
\usepackage{needspace}
\usetikzlibrary{arrows.meta}
\usepgfplotslibrary{groupplots}
\pgfplotsset{compat=1.18,every axis/.append style={font=\scriptsize,axis line style={black!65},tick style={black!65},label style={font=\scriptsize},title style={font=\scriptsize},scaled ticks=false}}
\definecolor{spinblue}{RGB}{47,94,161}
\definecolor{spinred}{RGB}{194,63,68}
\definecolor{spingreen}{RGB}{38,134,102}
\pgfplotsset{colormap={spinmap}{rgb255(0cm)=(47,94,161);rgb255(1cm)=(238,238,238);rgb255(2cm)=(194,63,68)}}
\newlength{\edgefigwidth}
\AtBeginDocument{\setlength{\edgefigwidth}{\columnwidth}\ifdim\edgefigwidth>85mm\setlength{\edgefigwidth}{85mm}\fi}
\newlength{\edgewidefigwidth}
\AtBeginDocument{\setlength{\edgewidefigwidth}{\textwidth}\ifdim\edgewidefigwidth>178mm\setlength{\edgewidefigwidth}{178mm}\fi}

\usepackage[colorlinks=true,allcolors=blue!50!black]{hyperref}
\renewcommand{\section}[1]{\par\Needspace{4\baselineskip}\medskip\noindent\textit{#1.}\enspace\ignorespaces}
\begin{document}
\title{Magnetically Compensated Spin-Polarized Edge States in Coplanar Magnets}
\author{Dinghui Wang}
\email{wangdh@cumt.edu.cn}
\affiliation{School of Materials Science and Physics, China University of Mining and Technology, Xuzhou 221116, China}
\affiliation{Jiangsu Physical Science Research Center, Nanjing 210093, China}
\author{Lulu Liu}
\affiliation{School of Electronic Engineering, Nanjing Xiaozhuang University, Nanjing, 211171, China}
\begin{abstract}
A boundary can release spin-polarized electronic states from a magnetically compensated, spin-degenerate layer. We establish the symmetry criterion governing this release in coplanar exchange magnets: the allowed even-parity edge-spin space equals the allowed uniform-magnetization space. Symmetry-protected compensation therefore selects an odd spin texture perpendicular to the moment plane. Six maximal sectors organize the possible edge textures. Orthogonal cuts of one exchange model realize odd and even sectors, while FeSe spin-vortex and NiI$_2$ period-four spiral Wannier models provide material realizations. Their semi-infinite boundaries exhibit complementary electrical responses: perpendicular spin accumulation at the odd edge and in-plane spin-current response at the even edge. Boundary symmetry thus connects magnetic compensation to equilibrium spin textures and electrical spin readout.
\end{abstract}
\maketitle
Spin-polarized electronic states can coexist with magnetic compensation when symmetry relates opposite polarizations at different momenta. Bulk altermagnets realize this combination through crystal rotations and even-parity spin textures.\cite{Smejkal2022a,Smejkal2022b} Coplanar noncollinear magnets additionally admit odd-parity spin perpendicular to the local-moment plane,\cite{Hellenes2023,Brekke2024} providing a nonrelativistic mechanism for electrically induced spin accumulation.\cite{Chakraborty2025} A boundary offers a geometric way to select between these possibilities: the exchange pattern remains fixed while the cut removes spatial operations that constrain the electronic spin.

Termination-induced polarization is established for collinear antiferromagnetic surfaces,\cite{Lange2026,Hu2026,Sattigeri2023} and coplanar kagome ribbons exhibit termination-dependent perpendicular spin.\cite{LyHayami2026} Orientation-selective spin-polarized states have also been observed at monolayer NiI$_2$ edges.\cite{NiI2Edge2024} These results motivate a distinct symmetry question for a single one-dimensional edge: which polarized textures have a uniform moment forbidden by the surviving symmetry group? An edge momentum can only be preserved or reversed. The angular rotations that compensate an even-wave texture in a two-dimensional Brillouin zone consequently have a different role after projection onto one edge.

The answer is the identity between the allowed even-parity edge-spin space and the allowed uniform-spin magnetization space. It selects odd perpendicular spin as the polarized sector with symmetry-protected compensation and organizes all possibilities into six maximal sectors. A scalar-hopping exchange model exposes the mechanism through two orthogonal cuts of one spin-degenerate parent. The same classification determines complementary dissipative readouts: odd spin produces accumulation, whereas even spin produces a conventional spin-current response. FeSe spin-vortex and NiI$_2$ period-four spiral half-spaces connect this sequence of symmetry, spectrum, and response to material Hamiltonians.

\begin{figure}[!tbp]
\centering
\includegraphics[width=\edgefigwidth]{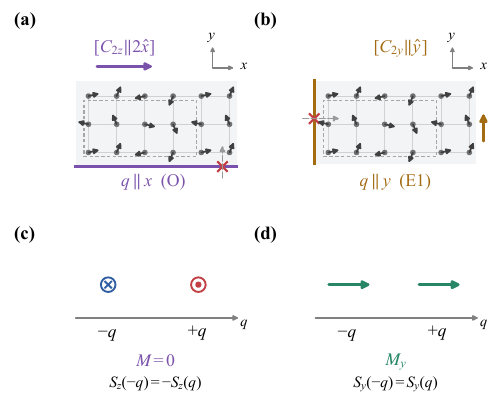}
\caption{Boundary selection of electronic spin from one coplanar exchange pattern. (a,b) Alternative cuts of the same model; gray arrows are local moments, dashed outlines mark a $4\times2$ magnetic cell, and colored lines mark the boundary. The tangent spin translation survives; the dashed arrow crossing the cut, marked by a red cross, is removed. The retained operations are $[C_{2z}\Vert2\hat x]$ for (a) and $[C_{2y}\Vert\hat y]$ for (b). (c,d) Schematic electronic spins at paired edge momenta. Dot/cross denote $+z/-z$, while green arrows denote $+y$. Odd perpendicular spin compensates exactly; equal even spins permit a uniform moment.}
\label{fig:concept}
\end{figure}

\section{Single-edge symmetry and magnetic compensation}

Consider a fixed exchange order with local moments in the spin $xy$ plane, real scalar hopping, and no spin--orbit coupling or orbital flux. A straight edge retains translations along its tangent and leaves a one-dimensional momentum $q=k_\parallel L_e$, where $L_e$ is the magnetic edge period. The surviving spin-space operations\cite{Brinkman1966,Litvin1974,Xiao2024} must preserve the same material half-plane. Writing an affine operation as $\mathbf r\mapsto R_g\mathbf r+\mathbf t_g+\mathbf L$, the condition for a primitive normal covector $\mathbf n$ is
\begin{equation}
 \mathbf n R_{g,2}=\mathbf n,\qquad
 \mathbf n\cdot(\mathbf t_g+\mathbf L)=0.
 \label{eq:halfplane-main}
\end{equation}
The integer lattice lift $\mathbf L$ is included before taking the edge point group. Tangent reflection can survive, whereas normal reversal exchanges material and vacuum. Figure~\ref{fig:concept}(a,b) illustrates the corresponding loss of a transverse spin translation.

For a complete invariant near-edge region, let $P_e$ select its orbitals. The physical Pauli-spin operator assigned to that region is $\Sigma_i=\{P_e,\sigma_i\}/2$, and its spectral density is
\begin{equation}
 A_i(q,E)=\operatorname{Tr}[\Sigma_i\delta(E-H(q))].
 \label{eq:spectral}
\end{equation}
The charge weight $A_0$ uses $P_e$ in place of $\Sigma_i$, and $P_i=A_i/A_0$ is the conditional spin polarization. The trace covers complete degenerate subspaces and all states contributing to the region, allowing the same definition to describe isolated edge bands and resonances coupled to the layer continuum.

Every such coplanar order retains the spin-only antiunitary operation $\Theta=C_{2z}\mathcal T$. Its physical-spin action is $M=\operatorname{diag}(1,1,-1)$, so
\begin{equation}
 \mathbf A(-q,E)=M\mathbf A(q,E).
 \label{eq:coplanar-main}
\end{equation}
Thus electronic in-plane spin is even in $q$ and perpendicular spin is odd. For any additional retained operation $g$, let $S_g$ act on physical spin and $a_g=\pm1$ on the real-space tangent. Its final momentum action is $\alpha_g=(\det S_g)a_g$, including the reversal from antiunitarity. Resolving the covariance $\mathbf A(\alpha_gq,E)=S_g\mathbf A(q,E)$ into even and odd parts gives
\begin{equation}
 V_p=\bigcap_{g\in G_{\mathbf n}}\ker(S_g-\alpha_g^p I),
 \qquad p=0,1.
 \label{eq:spaces}
\end{equation}
The separation is global in momentum: all even harmonics belong to $V_0$ and all odd harmonics to $V_1$.

Uniform spin magnetization has no momentum label and transforms only with $S_g$. Its allowed space is therefore
\begin{equation}
 V_{\mathcal M}=\bigcap_{g\in G_{\mathbf n}}\ker(S_g-I)=V_0.
 \label{eq:compensation}
\end{equation}
This identity makes compensation a constraint on the entire spectral texture. If symmetry forbids a uniform moment, every even component of $\mathbf A(q,E)$ vanishes. A polarized compensated edge then has only odd perpendicular spin. Integrating the odd spectrum with the equilibrium Fermi function cancels its electronic moment between $q$ and $-q$ at every filling and temperature [Fig.~\ref{fig:concept}(c)]. An allowed even moment instead depends on the microscopic spectrum and filling [Fig.~\ref{fig:concept}(d)].

\section{Six sectors from retained operations}

Equation~\eqref{eq:coplanar-main} bounds the dimensions by $d_e=\dim V_0\in\{0,1,2\}$ and $d_o=\dim V_1\in\{0,1\}$. Their $3\times2$ combinations give the six maximal sectors in Table~\ref{tab:sectors}. These sectors specify allowed spin spaces; the Hamiltonian fixes the amplitudes within them. In particular, mixed electronic textures can span three spin dimensions even though every exchange moment remains in one plane.

\begin{table}[tbp]
\small
\setlength{\tabcolsep}{3pt}
\caption{Complete maximal single-edge spin sectors. The in-plane axis in E1/M2 is chosen as $y$. The allowed uniform-magnetization space equals $V_0$; O is the polarized sector with protected compensation.}
\label{tab:sectors}
\begin{tabular}{lccc}
\hline\hline
Sector & $(d_e,d_o)$ & Even space $V_0$ & Odd space $V_1$\\
\hline
U & $(0,0)$ & $0$ & $0$\\
O & $(0,1)$ & $0$ & $z$\\
E1 & $(1,0)$ & $y$ & $0$\\
E2 & $(2,0)$ & $x,y$ & $0$\\
M2 & $(1,1)$ & $y$ & $z$\\
M3 & $(2,1)$ & $x,y$ & $z$\\
\hline\hline
\end{tabular}
\end{table}

Representative generators make the classification explicit. A tangent translation paired with $C_{2z}$ preserves $q$ but changes $(A_x,A_y,A_z)$ to $(-A_x,-A_y,A_z)$. It removes the even plane and leaves O. Pairing a tangent translation with $C_{2y}$ instead leaves only even $A_y$, giving E1. Their simultaneous retention in a complete invariant region removes all spin components and gives U; a two-sublayer model realizes the second constraint through $[C_{2y}\Vert m_z]$.

Tangent reflections supply the other cases. Let $m_q$ reverse the edge coordinate and preserve the material side. The operation $[I\Vert m_q]$ makes the entire spin spectrum even, retaining the two-dimensional in-plane space E2. The paired operation $[C_{2y}\Vert m_q]$ has $\alpha=-1$: its even equation removes $A_x$, while its odd equation retains $A_z$, producing M2. With only the common constraints effective, both even in-plane directions and the odd perpendicular direction remain, giving M3. Equation~\eqref{eq:spaces} classifies any chosen edge through its full retained subgroup.

Higher harmonics change the texture shape within these spaces. Near an invariant momentum an odd component has $A_z=c_1(E)q+c_3(E)q^3+\cdots$, with a linear $p$-wave form when $c_1\ne0$. All odd terms obey the same one-dimensional symmetry condition. The rational-edge construction applied to a 4257-entry coplanar spin layer-group catalog identifies 726 entries that forbid bulk spectral spin and admit an O edge. The Supplemental Material (SM) gives the catalog provenance, complete rational-direction coverage, and model realizations of all six sectors.

\section{Boundary release from a degenerate parent}

\begin{figure}[!tbp]
\centering
\includegraphics[width=\edgefigwidth]{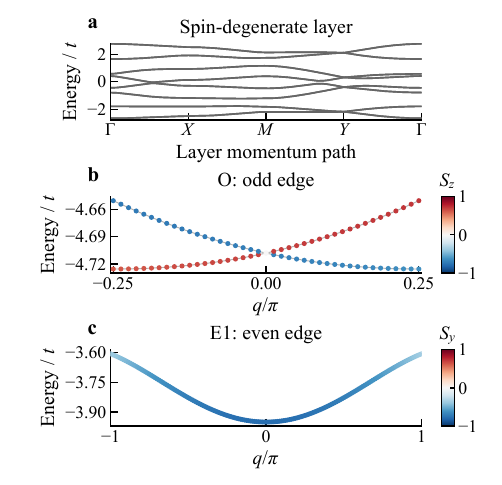}
\caption{Computed model spectra and their momentum geometry. (a) Path in the magnetic Brillouin zone, with $k_x,k_y$ the phases across the $4\times2$ magnetic cell. (b) All parent bands along that path. (c) Lowest O doublet near $q=0$, colored by physical $S_z$. (d) Lowest E1 branch, colored by $S_y$ with the same Pauli-spin scale. Degenerate multiplets use their averaged spin trace. Both cuts have $J=0.85t$, width 32, and scalar outer-row potential $V_e=-2.5t$; the full exchange texture is defined in Fig.~\ref{fig:concept} and Eq.~\eqref{eq:moment-main}.}
\label{fig:release}
\end{figure}

The two cuts in Fig.~\ref{fig:concept} share the microscopic Hamiltonian
\begin{equation}
 H=\sum_{ij}t_{ij}c_i^\dagger c_j+
 J\sum_i c_i^\dagger\mathbf m_i\cdot\boldsymbol\sigma c_i,
 \label{eq:model}
\end{equation}
with real scalar hopping and on-site terms. Its four-column, two-row exchange pattern is
\begin{equation}
 \begin{gathered}
 \mathbf m_{jl}=s_j\big((-1)^l\cos\theta_{j\bmod2},
                   \sin\theta_{j\bmod2},0\big),\\
 s_{j+2}=-s_j,
 \end{gathered}
 \label{eq:moment-main}
\end{equation}
where $s_0=s_1=1$ and $(\theta_0,\theta_1)=(0.23,1.17)$. Two-column translation carries $C_{2z}$, while one-row translation carries $C_{2y}$. The spinor operators commute with $H$ and anticommute with each other. Acting on an eigenstate of one operator, the other produces an orthogonal state at the same energy, enforcing the parent doublets. Their two spin constraints also force zero spin trace at each energy. Figure~\ref{fig:release}(a,b) defines the magnetic Brillouin-zone path and displays these bands.

An $x$ edge removes the row translation and retains the compensating $C_{2z}$ translation, releasing odd perpendicular O spin. The orthogonal edge retains $C_{2y}$ and releases even $y$-polarized E1 spin [Fig.~\ref{fig:release}(c,d)]. Rotating the cut switches the polarization from O to E1 while keeping the magnetic Hamiltonian fixed. A scalar outer-row potential moves the branches into an isolated spectral window; direct truncation retains the O sector as well (SM).

Projection onto the O doublet near $q=0$ gives
\begin{equation}
 \begin{gathered}
 H_e(q)=\epsilon(q)I+vq\mu_z+O(q^3),\\
 P\sigma_zP=s_0\mu_z+O(q),
 \end{gathered}
 \label{eq:edgeham}
\end{equation}
with $v=0.04679t$ and $s_0=0.68631$. The retained antiunitary half translation squares to $-e^{-iq}$ and protects the clean doublet at $q=0$. A local scalar defect breaks that translation and has a finite backscattering matrix element. The SM also constructs E2, M2, M3 and U by choosing tangent reflection, a coplanar outer-row field, or a complete two-sublayer geometry.

\section{Spin parity as an electrical selection rule}

A longitudinal field redistributes the occupations of opposite-moving states. For electron charge $q_e=-e$ and relaxation time $\tau$,
\begin{equation}
 \delta f_n=q_eE_\parallel\tau v_n^{\rm phys}
                    (-\partial_E f_0).
 \label{eq:distribution-main}
\end{equation}
Equal-energy partners at $q,-q$ have opposite velocities and opposite $\delta f$. Odd spin therefore adds in the induced density, whereas even spin cancels. A convective spin flux contributes one more velocity factor, reversing this selection: its field-induced response adds for even spin.

In model units $\varepsilon=E/t$, $\theta=k_BT/t$, $v_n=\partial_q\varepsilon_n$ and $K_n=-\partial_{\varepsilon_n}f_0$, the corresponding kernels are
\begin{equation}
 \begin{aligned}
 D&=\sum_n\int\frac{dq}{2\pi}v_n^2K_n,\\
 \mathbf X&=\sum_n\int\frac{dq}{2\pi}\mathbf s_n v_nK_n\ \in V_1,\\
 \mathbf B&=\sum_n\int\frac{dq}{2\pi}\mathbf s_n v_n^2K_n\ \in V_0.
 \end{aligned}
 \label{eq:occupation-main}
\end{equation}
Here $\mathbf s_n$ is physical Pauli spin; complete degenerate subspaces are treated by their velocity-resolved traces. $D$ measures charge transport, $\mathbf X$ spin accumulation, and $\mathbf B$ the convective spin-flux response. For $u=(\tau t/\hbar)(-eE_\parallel L_e/t)$, the physical responses are $I=-(et/\hbar)uD$, $\delta\mathbf S=(\hbar/2)u\mathbf X$, and $\delta\mathbf J^{\rm conv}=(t/2)u\mathbf B$.

\begin{figure}[!tbp]
\centering
\includegraphics[width=\edgefigwidth]{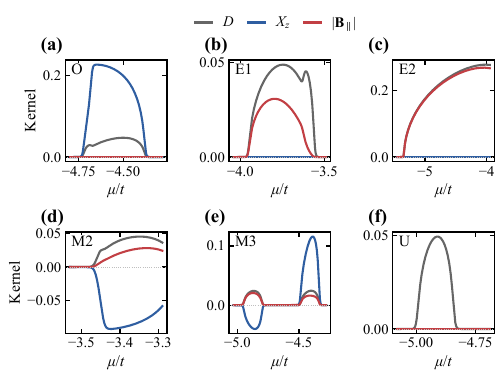}
\caption{Computed occupation transport across the six microscopic edge sectors: (a) O, (b) E1, (c) E2, (d) M2, (e) M3, and (f) U. Gray curves give the charge kernel $D$, blue curves the perpendicular density kernel $X_z$, and red curves the in-plane convective flux magnitude $|\mathbf B_\parallel|$. All use Eq.~\eqref{eq:occupation-main}, $k_BT=0.004t$, and isolated edge-band windows. The density reads odd spin and the flux reads even spin; mixed sectors support both. U remains conducting while both occupation spin kernels cancel. Each window follows its own model dispersion; full constructions are in the SM.}
\label{fig:model-transport}
\end{figure}

Figure~\ref{fig:model-transport} tests this selection in all six models. O supports perpendicular accumulation; E1 and E2 support respectively one- and two-axis in-plane flux. M2 and M3 combine the odd density with the even flux, while U retains a finite charge channel. The dispersion controls the relative signal size. At $\mu/t=-4.64747$, the O model has $X_z=0.22635$ and $D=0.0322875$, giving $X_z/D\simeq7.01$ in these density-to-flux units. A slow strongly polarized branch can thus carry appreciable spin accumulation per charge current.

The conventional spin current accounts for spin--velocity noncommutation through $j_i=\{v,O_i\}/2$, in place of the product of band averages. Its occupation response has the same allowed space $V_0$, with an amplitude determined by this operator. Coherent interband responses have a different antiunitary selection: complex conjugation changes the sign of the imaginary matrix-element product. Their density and current spaces are $W_S=\bigcap_g\ker(S_g-a_gI)$ and $W_J=\bigcap_g\ker(S_g-\det S_g I)$. For example, M2 allows coherent density perpendicular to its static $y,z$ texture, and a representative U$_C$ model supports coherent in-plane spin generation. The SM develops these interband, thermal, and finite-bias extensions. The material comparison below evaluates the dissipative regional density and conventional current with one consistent operator definition.

\section{Magnetic structure and spectra of material edges}

\begin{figure*}[t]
\centering
\includegraphics[width=\edgewidefigwidth]{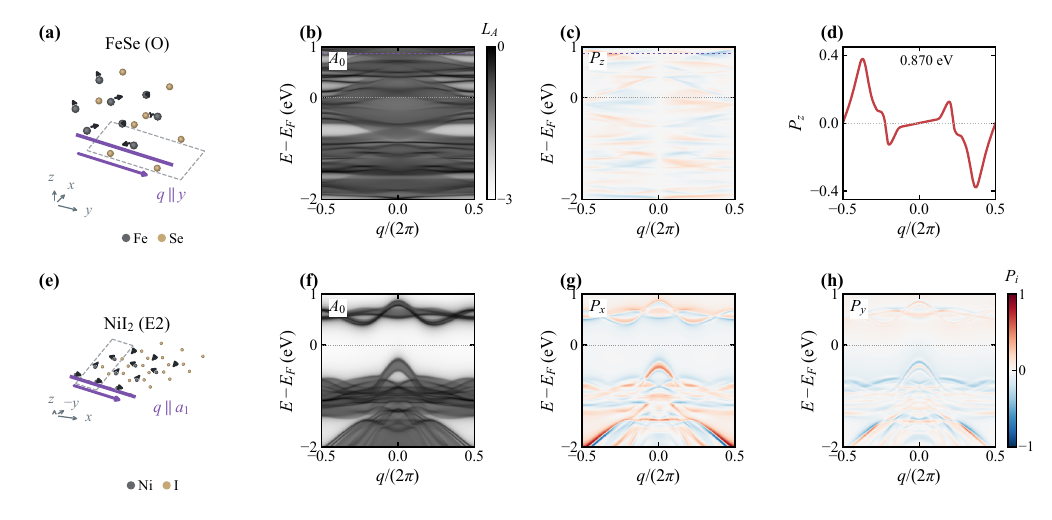}
\caption{Material geometry and computed single-boundary spectra. Rows compare FeSe and NiI$_2$. (a,e) Three-dimensional views from the actual fixed-ion QE coordinates, with metal sites gray, Se/I gold, and target metal moments shown by dark arrows. Purple lines and arrows define the ideal edge and tangent; dashed outlines mark one magnetic edge cell. Local triads give the physical axes; edge tangents are $y$ for FeSe and $x\parallel\mathbf a_1$ for NiI$_2$. The cuts retain complete magnetic cells. (b,f) Charge weight $L_A=\log_{10}(A_0/\max A_0)$. (c,g,h) Physical Pauli polarizations $P_z$ in FeSe and $P_x,P_y$ in NiI$_2$, retaining original Cartesian spin axes. (d) FeSe momentum cut at $E-E_F=0.870$~eV, marked by purple guides in (b,c). Maps retain $201\times601$ points, half-width 10~meV, and a region of one complete magnetic Wannier cell.}
\label{fig:materials}
\end{figure*}

The FeSe spin-vortex phase\cite{OHalloran2017,Shishidou2018} realizes the degenerate-parent mechanism in a metal--ligand layer [Fig.~\ref{fig:materials}(a)]. Its nearest-neighbor Fe directions are $\mathbf e_1=(\mathbf a+\mathbf b)/2$ and $\mathbf e_2=(-\mathbf a+\mathbf b)/2$, where $\mathbf a,\mathbf b$ are the chemical square axes. The prescribed moments satisfy $\mathbf m_{ij}=(2\mu_B/\sqrt2)((-1)^i,(-1)^j,0)$. Translation along either Fe direction exchanges upper and lower Se environments and pairs with $m_z$ and a spin rotation, $C_{2y}$ or $C_{2x}$. The resulting spinor glides anticommute, enforcing parent doublets. The fitted model retains doublets within 0.24~meV on 441 sampled layer momenta.

A chemical-$y$ edge removes the two transverse glides but retains their tangent combination with spin action $C_{2z}$. It therefore implements the O constraint, with predominantly odd $P_z$ reaching about 0.38 [Fig.~\ref{fig:materials}(b--d)]. The charge map locates the polarization in the complete regional spectrum, and the common energy guide connects that map to its odd momentum cut. The nearest-neighbor Fe directions differ from the chemical axes: their corresponding edges can retain a different glide and realize E1.

The second material is a period-four spiral approximant of monolayer NiI$_2$,\cite{NiI2Mono2022,NiI2Switch2025} with target moments $\mathbf m_j=2\mu_B(\cos(\pi j/2),\sin(\pi j/2),0)$ on consecutive Ni rows [Fig.~\ref{fig:materials}(e)]. The $90^\circ$ coplanar sequence forms a commensurate spiral. The $\mathbf a_1$ cut retains tangent reflection with identity spin action. It removes the odd normal component and gives E2, with separate in-plane polarization maxima $|P_x|\simeq0.86$ and $|P_y|\simeq0.53$ [Fig.~\ref{fig:materials}(f--h)]. The two material examples implement complementary retained operations: a compensating spin translation and a spin-preserving tangent reflection.

The semi-infinite construction keeps one physical boundary and couples the selected region to the entire interior. Spinor Wannier models retain metal $d$ and ligand $p$ orbitals, including physical spin and the position connection in the Hamiltonian gauge. The regional trace consequently includes both edge resonances and the surrounding layer spectral weight. This connects the ideal-cut symmetry analysis directly to the observable in Eq.~\eqref{eq:spectral}.

\section{Complementary material spin responses}

\begin{figure*}[t]
\centering
\includegraphics[width=\edgewidefigwidth]{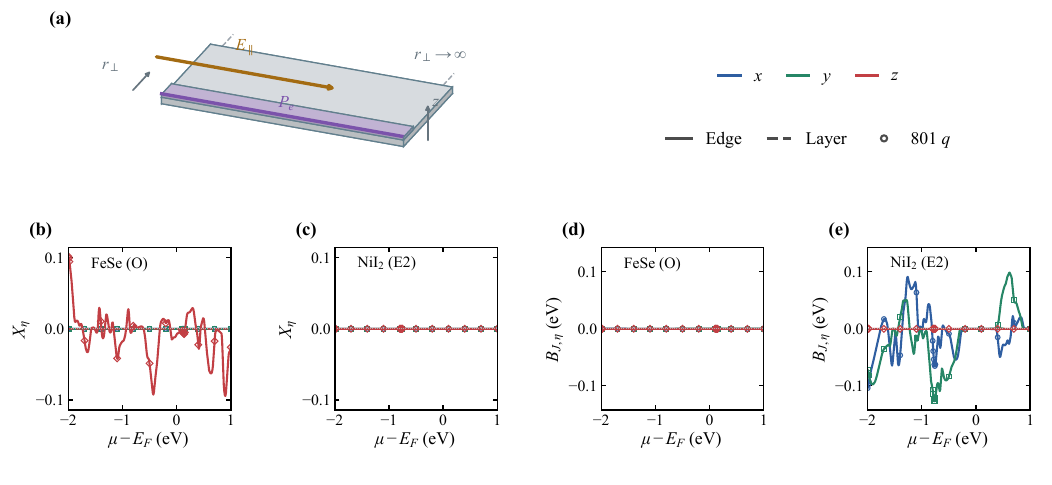}
\caption{Geometry and computed dissipative readout of one edge. (a) Three-dimensional schematic of the uniform longitudinal field and regional projector $P_e$ beside the edge. The marked normal $z$ and inward coordinate $r_\perp$ distinguish spin and propagation directions; dashed continuation represents the semi-infinite interior. (b,c) Density kernels $X_{\eta,i}=2\eta K_{\Sigma_i}^e/L_e$. (d,e) Conventional Pauli-current kernels $B_{J,\eta,i}=2\eta K_{\mathcal J_i}^e/L_e^2$, in eV. Panels (b,d) show FeSe and (c,e) show NiI$_2$. Blue/green/red give $x/y/z$. Solid lines: edge region; dashed: periodic-layer controls; open circles/squares/diamonds: independent 801-$q$ checks for $x/y/z$. Curves retain 301 chemical potentials and 201 midpoint momenta with common response scales, $\eta=20$~meV, and $T=0$.}
\label{fig:response}
\end{figure*}

A uniform longitudinal field probes the same near-edge region [Fig.~\ref{fig:response}(a)]. Let $W=\hbar v_\parallel$ and $\delta_\eta=(G^A-G^R)/(2\pi i)$ at chemical potential $\mu$. The dissipative half-space kernel for a regional observable $O_e$ is
\begin{equation}
 K_O^e(\mu)=\pi\int\frac{dq}{2\pi}
          \operatorname{Re}\operatorname{Tr}[O_e\delta_\eta W\delta_\eta].
 \label{eq:kubo}
\end{equation}
Accumulation uses $O_e=\Sigma_i$, while the conventional regional Pauli current uses $\mathcal J_i=\{P_e,\{\sigma_i,W\}/2\}/2$. The velocity includes the Hermitian position connection and its commutator after cutting the Hamiltonian. This retains the boundary contribution to the physical current vertex.

A density response contains one velocity factor, whereas a current response contains two. Their transformation under $q\mapsto\alpha_gq$ therefore gives the same complementary spaces as the mode argument:
\begin{equation}
 \mathbf K_\Sigma^e\in V_1,\qquad
 \mathbf K_{\mathcal J}^e\in V_0.
 \label{eq:response}
\end{equation}
O realizes an edge Edelstein response with perpendicular spin accumulation.\cite{Chakraborty2025,Gonzalez2024} E1/E2 select an in-plane conventional current response. This statement applies to the regional half-space contraction, including hybridization with the continuum.

The computed curves show this contrast directly. FeSe has a leading perpendicular accumulation peak $|X_{\eta,z}|\simeq0.102$ [Fig.~\ref{fig:response}(b)]. NiI$_2$ has in-plane current components with separate maxima $|B_{J,\eta,x}|\simeq0.104$ and $|B_{J,\eta,y}|\simeq0.127$~eV [Fig.~\ref{fig:response}(e)]. Their relative sign and direction depend on filling, while the permitted plane remains fixed. Complementary components lie near the numerical background, and periodic-layer controls are substantially smaller. Independent 801-momentum checks change the leading NiI$_2$ components by less than 0.5\%; FeSe agrees to the plotted precision. The response therefore follows the symmetry-selected channel across a continuous chemical-potential range.

The plotted normalization separates the common scalar lifetime from the spectral kernels: $X_\eta=2\eta K_\Sigma^e/L_e$ and $B_{J,\eta}=2\eta K_{\mathcal J}^e/L_e^2$. The chosen $\eta=20$~meV corresponds to $\tau=\hbar/(2\eta)=16.455$~fs. For electron force $F=-eE_\parallel$, the angular-momentum density per unit edge length is $\delta s_i=(\hbar/2)F K_{\Sigma_i}^e/L_e$. Since the current vertex uses $W=\hbar v_\parallel$, the angular-momentum flux is $\delta j_i=F K_{\mathcal J_i}^e/(2L_e)$. These conversions specify the density and conventional-current observables accessed by edge-sensitive detection under bias.

\paragraph{Computational setting.}
Constrained noncollinear QE calculations\cite{Giannozzi2009} evaluate fixed-ion FeSe spin-vortex and NiI$_2$ period-four phases, using LDA and LDA+$U_{\rm eff}=1$~eV, respectively. Spinor Wannier models\cite{Pizzi2020} contain 64 and 88 orbitals. Independent QE comparisons at four additional momenta give maximum energy deviations of 2.57/3.01~meV and spin-cluster eigenvalue deviations of 0.0352/0.0354 in Pauli units for FeSe/NiI$_2$. Complete metal--ligand motifs define the ideal cuts. Semi-infinite Green functions\cite{Wu2018,LopezSancho1985} and a mixed-resolvent embedding evaluate the spectra and Eq.~\eqref{eq:kubo}. The SM specifies atomic geometry, magnetic constraints, projector assignment, operator construction, and convergence controls.

Boundary geometry controls both the symmetry that forbids a moment and the electronic channel through which spin is read. The identity $V_{\mathcal M}=V_0$ connects these two roles: protected compensation removes all even spin, leaving an odd perpendicular texture and its accumulation response. Retaining an even sector instead permits an in-plane current response, and mixed sectors combine both. The common-parent model isolates the effect of the cut, while the two material phases carry the same logic into orbital-resolved half-spaces. A reconstruction or coplanar boundary field can change the retained subgroup and move between sectors; its outcome follows by repeating the same pair of kernel intersections. Spin-resolved edge spectroscopy and local spin detection under current probe the equilibrium texture and its driven counterpart.
\begin{acknowledgments}
D. Wang is supported by Natural Science Foundation of Jiangsu Province (No. BK20233001). L. Liu is supported by the Basic Research Program of Jiangsu (Grant SBK20260202315).
\end{acknowledgments}
\paragraph*{Supporting Information}
Single-edge subgroup and rational-direction construction; complete classification counts and six-sector model realizations; direct-cut localization and scalar-defect control; constrained density-functional and Wannier methods and independent energy/spin accuracy; semi-infinite spectral and mixed-resolvent response methods, full charge/periodic controls and integration checks; model occupation, coherent, thermal, nonlinear and ballistic responses.
\paragraph*{Data Availability}
The accompanying \texttt{research\_data.zip} contains the symmetry database, edge catalogs, material Wannier operators, spectra, response arrays, structures, and analysis code. Database provenance is given in the Supporting Information.

\end{document}